\documentclass{article}\usepackage[]{graphicx}\usepackage[]{xcolor}
\makeatletter
\def\maxwidth{ %
  \ifdim\Gin@nat@width>\linewidth
    \linewidth
  \else
    \Gin@nat@width
  \fi
}
\makeatother

\definecolor{fgcolor}{rgb}{0.345, 0.345, 0.345}

\usepackage{framed}
\makeatletter
 {\par\unskip\endMakeFramed%
 \at@end@of@kframe}
\makeatother

\definecolor{shadecolor}{rgb}{.97, .97, .97}
\definecolor{messagecolor}{rgb}{0, 0, 0}
\definecolor{warningcolor}{rgb}{1, 0, 1}
\definecolor{errorcolor}{rgb}{1, 0, 0}

\usepackage{alltt}

\usepackage[default]{sourcesanspro} 
\usepackage[english]{babel}
\usepackage[utf8]{inputenc}
\usepackage{amsmath}        
\usepackage{natbib}
\usepackage{url}
\usepackage{todonotes}
\usepackage{bbold}          
\usepackage{booktabs}
\usepackage{setspace}       

\makeatletter
\let\start@align@nopar\start@align
\let\start@gather@nopar\start@gather
\let\start@multline@nopar\start@multline
\long\def\start@align{\par\start@align@nopar}
\long\def\start@gather{\par\start@gather@nopar}
\long\def\start@multline{\par\start@multline@nopar}
\makeatother
\definecolor{linkblue}{RGB}{19,168,158}
\definecolor{linkgreen}{RGB}{27,183,117}

\usepackage{hyperref}
\usepackage{href-ul}
\renewcommand\url[1]{{\href{#1}{#1}}}
\hypersetup{
    colorlinks,
    citecolor=linkgreen,
    filecolor=black,
    linkcolor=linkgreen,
    urlcolor=linkgreen,
    }

\usepackage{geometry} 
\newcommand{\given}{\,\vert\,}

\newcommand\independent{\protect\mathpalette{\protect\independenT}{\perp}}
\def\independenT#1#2{\mathrel{\rlap{$#1#2$}\mkern2mu{#1#2}}}

\title{Outcomes truncated by death in RCTs: a simulation study on the survivor average causal effect}
\IfFileExists{upquote.sty}{\usepackage{upquote}}{}
\begin{document}

\maketitle

\noindent
Stefanie von Felten$^{1,*}$, Chiara Vanetta$^{1,2}$, Christoph M. Rüegger$^{3}$, Sven Wellmann$^{4}$ \& Leonhard Held$^{1}$\\

\noindent
$^1$Department of Biostatistics at Epidemiology, Biostatistics and Prevention Institute, University of Zurich, Hirschengraben 84, CH-8001 Zurich, Switzerland\\
$^2$Biostatistics and Research Decision Sciences, MSD, The Circle 66, CH-8058 Zurich, Switzerland\\
$^3$Newborn Research, Department of Neonatology, University Hospital Zurich, University of Zurich, Frauenklinikstrasse 10, CH-8091 Zürich, Switzerland\\
$^4$Department of Neonatology, University Children’s Hospital Regensburg, Hospital St Hedwig of the Order of St John, University of Regensburg, Steinmetzstraße 1-3, D-93049 Regensburg, Germany\\
$^*$Corresponding author: e-mail: stefanie.vonfelten@uzh.ch, Phone: +41-44-634-46-44


\section*{Abstract} 
Continuous outcome measurements truncated by death present a challenge for the estimation of unbiased treatment effects in randomized controlled trials (RCTs). 
One way to deal with such situations is to estimate the survivor average causal effect (SACE), but this requires making non-testable assumptions.
%
Motivated by an ongoing RCT in very preterm infants with intraventricular hemorrhage, we performed a simulation study to compare a SACE estimator with complete case analysis (CCA) and analysis after multiple imputation of missing outcomes. 
We set up 9 scenarios combining positive, negative and no treatment effect on the outcome (cognitive development) and on survival at 2 years of age.
Treatment effect estimates from all methods were compared in terms of bias, mean squared error and coverage with regard to two true treatment effects: the treatment effect on the outcome used in the simulation and the SACE, which was derived by simulation of both potential outcomes per patient.
%
Despite targeting different estimands (principal stratum estimand, hypothetical estimand), the SACE-estimator and multiple imputation gave similar estimates of the treatment effect and efficiently reduced the bias compared to CCA. Also, both methods were relatively robust to omission of one covariate in the analysis, and thus violation of relevant assumptions.
%
Although the SACE is not without controversy, we find it useful if mortality is inherent to the study population.
Some degree of violation of the required assumptions is almost certain, but may be acceptable in practice.

\paragraph{Keywords:} Estimand; Multiple Imputation; Principal Stratification; SACE

\section{Introduction}

Analysis by the intention-to-treat (ITT) principle was recommended for the primary analysis of randomized clinical trials (RCTs) by the \cite{ICH_E9} guideline and is still regarded the gold standard for most RCTs.
Analysis by ITT, including all randomized patients and analyzing them by the randomized treatment, irrespective of whether this treatment was received exactly as planned, best preserves the benefits of randomization and estimates the effect of a ``treatment policy''.
However, the latter may not always represent the most relevant ``estimand'', and this is addressed in the \cite{ICH_E9_addendum} addendum.
While the addendum undisputably includes the treatment policy strategy (conform with ITT), it discusses several alternative strategies to choose an estimand in the light of ``intercurrent events''.
Intercurrent events occur between randomization and outcome, and thus affect the interpretation or existence of the latter.
Examples include switching to a different treatment, treatment discontinuation, or death.
Although the addendum is fully agnostic of any techniques of estimation, estimation aligned with some strategies mentioned in the addendum (namely hypothetical and principal stratum strategies) requires ``causal inference'' techniques involving potential outcomes \citep{Imbens2015, Hernan2020}.
While these techniques were originally designed to draw causal inference from observational data, they have a role in RCTs when we can no longer compare like with like due to the presence of intercurrent events.

The most drastic intercurrent event in a trial is a patient's death. 
The event of death is a terminal event, for which the \cite{ICH_E9_addendum} addendum states that \emph{``in general, the treatment policy strategy cannot be implemented for intercurrent events that are terminal events, since values for the variable after the intercurrent event do not exist. For example, an estimand based on this strategy cannot be constructed with respect to a variable that cannot be measured due to death''}.
In an RCT with a time-to-event outcome death may be treated as a competing risk or it may be included in the primary outcome using a ``composite variable strategy''.
However, estimation of a causal treatment effect with death as intercurrent event is more challenging in an RCT with a continuous outcome, in particular if this variable is assessed only once.

An example of such an RCT is the EpoRepair trial, an ongoing placebo-controlled, parallel group, double-blind RCT on the effect of erythropoietin (Epo) for the repair of cerebral injury in very preterm infants suffering from intraventricular hemorrhage \citep{Ruegger2015, Wellmann2022}.
The primary outcome is the composite intelligence quotient (IQ) at five years of age measured using the Kaufmann Assessment Battery for Children (K-ABC). 
Cognitive development at two years of age, measured using the subscore for cognition of the Bayley Scales of Infant and Toddler Development (BSID-III), is a secondary outcome.
Both, IQ and cognitive development are measured only once.
The EpoRepair trial randomized 121 preterm infants to Epo or Placebo in a 1:1 ratio. However, 19 (15.7\,\%) of these vulnerable infants died before the age of two years (15 died before term equivalent age).
Further, a group blinded interim safety analysis indicated an imbalance in mortality between trial arms.
How should the effect of the intervention (Epo) on IQ or on cognitive development be estimated in this case, or more conceptually, which estimand should be targeted, when a considerable proportion of outcome measurements is truncated by death?

One commonly used ad hoc solution to the problem is to restrict the analysis to the survivors (if all survivors have a measurement of the outcome) or more generally, to complete cases.
In fact, the use of complete case analysis in RCTs with missing data is prevalent \citep{Zhang2017, Ren2022}, although it is well-known that this leads to biased estimates of the treatment effect, e.g. with regard to a treatment policy estimand, as it corresponds to a nonrandomized comparison \citep[see for example][]{CHMP2010, Little2012}.
No bias would occur only when outcome measurements are ``missing completely at random'', i.e., the missingness is neither related to the assigned treatment nor to any characteristics of the patients.
The size of the bias depends on the amount of missing or truncated observations and on their relation with the assigned treatment. 
Such a relationship with the patients response to the treatment is often likely, and differing proportions of missing observations between trial arms may add to this concern \citep{Vickers2013}.
In situations when outcome measurements are not ``missing completely at random'', the treatment effect estimated by complete case analysis usually does not relate to a clear or relevant clinical question. 
This implies that it will also be biased with regard to estimands other than the treatment policy estimand.

Another ad hoc solution may be to treat the truncated outcomes as missing outcome measurements and use multiple imputation to recover the missing information.
The underlying assumption of multiple imputation is that measurements are \emph{missing at random}, which means that the missingness can be explained by observed data.
Multiple imputation has been recommended for analyzing trials with missing data (but recommendations are not restricted to trials) and is known to be a better choice than single imputation methods \citep{White2011, Groenwold2012, Vickers2013}.
A strength of multiple imputation is that the uncertainty about the imputations can be taken into account using Rubin’s rules \citep{Rubin1987} to combine the estimates from the imputed data sets. 
With the availability of software to generate imputations and to combine effect size estimates, for instance in the {\tt{R}} packages \texttt{mice} \citep{vanBuuren2011} or \texttt{Amelia} \citep{Honaker2011} this method has become increasingly popular.
Multiple imputation is certainly a good choice for imputing missing outcomes if they would have been observable.
However, outcome measurements truncated by death are not just a missing data problem: they are not defined (the IQ of a child who died before the IQ could be measured is not defined).
While using multiple imputation allows to analyze all randomized patients (including those who died) the estimated causal treatment effect is ``hypothetical'', and for a scenario in which no patient would have died.
Such a ``hypothetical strategy'' is usually more meaningful if the intercurrent event (unlike death) could be avoided by design in a future trial.
Nevertheless, in practice multiple imputation may sometimes be used to analyze trials with outcomes truncated by death, especially if truncation by death accounts for only part of the missing outcomes.
Multiple imputation may then also be used to impute missing outcomes of survivors only, which may be of limited use if some observations are truncated by death.
For the EpoRepair trial, it was planned that multiple imputation will be used to adjust for missing outcomes due to death or dropout \citep{Ruegger2015}.
We therefore focus on multiple imputation to address a hypothetical estimand, although there are alternatives like inverse probability weighting or G-computation \citep[see][who are linking these methods]{OlarteParra2023}.

An alternative causal effect of the treatment could be estimated using principal stratification \citep{Robins1986, Frangakis2002} to estimate the survivor average causal effect \citep[SACE, introduced by][]{Rubin1998, Rubin2006}, i.e., the treatment effect in those patients who would have survived under both treatments.
A related concept is the complier average causal effect \citep[CACE,][]{Angrist1996}.
Principal stratification is based on stratifying patients conditional on potential outcomes of a post-randomization variable, such as the event of death. 
The focus is then either on the "principal stratum" in which an event would occur on all treatments or would not occur on any of the treatments.
In the case of the SACE, the focus is on the principal stratum of "always survivors", those patients who would not have died under any of the treatments.
Principal stratum strategies are one of the five strategies mentioned in the ICH E9(R1) addendum on estimands and sensitivity analysis in RCTs \citep{ICH_E9_addendum}, which has fostered a wave of interest from clinical trialists in the theory and application of estimands and estimators based on principal stratification in recent years:  
\cite{Bornkamp2021} discuss the role of principal stratum estimands in drug development, give examples of research questions that may be addressed and give an overview of assumptions required for estimation; 
\cite{Lipkovich2022} provide a comprehensive tutorial on using principal stratification in the analysis of clinical trials.
However, since the principal strata are not observable, determining the patients who belong to the principal stratum of interest and estimating principal stratum effects rely on strong, untestable assumptions.
Hence, principal stratification methods were strongly criticized \citep{Pearl2011, Joffe2011, Stensrud2023}.
Some authors have even doubted the existence of the principal stratum, or suspected that if it does exist, it may constitute a highly unusual subset of the population \citep{Dawid2012, Robins2007, Stensrud2023}.

Despite these reservations, we believe that the SACE is a useful estimand in the setting of the EpoRepair trial, for several reasons. 
First, the principal stratum strategy should be more relevant than the hypothetical strategy if an intercurrent event in a specific population cannot be avoided by design, as it is unfortunately the case with death among very preterm infants (and in other vulnerable or severely ill populations).
Second, while the majority of infants was expected to survive past 2 years, no treatment effect on survival was expected \emph{a priori}. As a consequence, we can expect the majority of the survivors to be always survivors (so we expect the principal stratum exists and even includes the majority of infants).

There is still limited awareness of the fact that outcomes truncated by death are not missing data in the usual sense. 
Further, if truncation by death is recognized as an issue, the choice of the estimand remains challenging.
We therefore performed a simulation study to compare complete case analysis (known to be biased in most cases), analysis after multiple imputation (targeting a hypothetical estimand), and analysis using a specific estimator of the SACE proposed by \cite{Hayden2005}.
The three methods are compared across 9 scenarios combining positive, negative and no treatment effect on the outcome and on survival with regard to bias, mean squared error and coverage.
Two types of true effects are considered: 1) the treatment effect on the outcome used in the simulation and 2) the SACE derived from the simulated observed and counterfactual data. 
Due to the strong assumptions underlying the chosen SACE estimator \citep{Hayden2005}, we present additional SACE methods with weaker assumptions \citep{Zhang2003, Chiba2011} in the supplementary material.
The EpoRepair trial was used as motivating example, and the simulated data resemble those from this trial.
With our work we wish to promote awareness of the issue of outcomes truncated by death among applied statisticans and clinicians, and methodological knowledge of how it could be dealt with.

\section{Methods}

We planned our simulation study by writing a simulation study protocol in advance (available on \url{https://osf.io/zafw3}), following the recommendations of \cite{Burton2006}.
For the description of the methods here we adopted the ADEMP structure (see Sections \ref{sec:m.aims}--\ref{sec:m.perf}) proposed for planning and reporting of simulation studies by \cite{Morris2019}.
Due to the complexity of the SACE estimation and the assumptions required, Section \ref{sec:methods.eval} contains detailed information on the specific SACE estimator used.
We consider our simulation study as a "neutral comparison study" \citep{Boulesteix2013}, because we compare three existing methods (rather than a new method with existing methods), which we have not developed ourselves, using evaluation criteria (bias, mean squared error and coverage) that were chosen in a rational way and were defined in the simulation study protocol in advance.

\subsection{Aims}\label{sec:m.aims}
The aim of our simulation study is to evaluate the performance of three different methods to estimate the treatment effect from a parallel group RCT when the primary outcome is truncated by death for a relevant proportion of the patients randomized.
We thereby wish to illustrate the challenge of estimating causal effects with outcomes truncated by death with a focus on applied statisticians and clinicians in clinical research as target readers.

\subsection{Data-generating mechanisms}\label{sec:m.dgm}

We simulated data from an RCT similar to the ongoing, placebo-controlled, double-blind EpoRepair trial on the effect of erythropoietin for the repair of cerebral injury in very preterm infants suffering from intraventricular hemorrhage \citep{Ruegger2015, Wellmann2022}.
Available data from EpoRepair were used as a basis, but we modified certain aspects to create the different scenarios for our simulation study. 
The outcome for this simulation study is cognitive development at two years of age (a secondary outcome in EpoRepair), measured using the subscore for cognition of the Bayley Scales of Infant and Toddler Development (BSID-III), hereafter referred to as outcome. 
We included gestational age (days), head circumference at birth (cm), socioeconomic status \citep[SES, ordinal, score from 2 to 12,][]{Largo1989} and Apgar score five min after birth (ordinal, from 1 to 10) as covariates $X_1$ to $X_4$.

We simulated data sets with a sample size of $n$=500 patients.
This sample size is similar to the size of a former Epo trial \citep{Natalucci2016}, but larger than the size of the EpoRepair trial for which only 121 patients were included \citep{Wellmann2022}.
We simulated the Apgar score based on the distribution observed in EpoRepair. 
We then simulated gestational age and head circumference as multivariate normal variables in each Apgar score category, using the means of these variables observed in each Apgar category with a constant variance-covariance matrix (estimated from EpoRepair) over all categories to avoid the simulation being too data-driven.
SES was simulated based on its observed distribution within Apgar score categories.
The patients were then randomly assigned to one of two treatments $z=0,1$; 250 patients received the Placebo control ($z=0$) and 250 received the intervention Epo ($z=1$).

The outcome $Y$ was simulated using the regression coefficients $b_1$, $b_2$ and $b_3$ of the covariates gestational age, head circumference and SES ($X_1$, $X_2$, and $X_3$) on the outcome, as estimated from the EpoRepair data, and a treatment effect $b_z$ on the outcome (mean difference, MD, Epo vs. Placebo), which we varied between simulation scenarios as $-5$, $0$ or $5$.
Survival $S$ was simulated using the regression coefficients $\tilde{b}_1$, $\tilde{b}_2$ and $\tilde{b}_4$ of the covariates gestational age, head circumference and Apgar score ($X_1$, $X_2$, and $X_4$) on survival as estimated from the EpoRepair data, and a treatment effect $\tilde{b}_z$  on survival (log odds ratio Epo vs. Placebo), which we varied again between simulation scenarios as -0.693, 0, or 0.693 (corresponding to odds ratios, OR, of 0.5, 1 and 2, respectively)\footnote{Note that \cite{Hayden2005} use $D$ for the survival status, which we found confusing, since $D$ may be associated with "death".}.
With this simulation procedure we aimed at an overall survival probability around 84\,\%, similar to the EpoRepair trial.

In addition to simulating data that could be observed, i.e., data under the allocated (and received) treatment, we simulated counterfactual data.
Using the same settings as described above, we simulated outcome and survival under the corresponding other treatment for each patient.
So, for each patient $i$, we simulated the outcome under treatment, $Y_i(z=1)$, and control, $Y_i(z=0)$:

\begin{align}
Y_i(1) &\sim N(b_0 + b_1\, X_{1,i} + b_2\, X_{2,i} + b_3 \,X_{3,i} + b_z, \sigma^2)\\
Y_i(0) &\sim N(b_0 + b_1\, X_{1,i} + b_2\, X_{2,i} + b_3 \,X_{3,i}, \sigma^2)
\end{align}

and then survival under treatment, $S_i(z=1)$, and control, $S_i(z=0)$:

\begin{align}
S_i(1) &\sim \text{Bernoulli}(p_i(1)), \quad \quad p_i(1) = \text{logit}^{-1}(\tilde{b}_0 + \tilde{b}_1\, X_{1,i} + \tilde{b}_2\, X_{2,i} + \tilde{b}_4 \,X_{4,i} + \tilde{b}_z)\\
S_i(0) &\sim \text{Bernoulli}(p_i(0)), \quad \quad p_i(0) = \text{logit}^{-1}(\tilde{b}_0 + \tilde{b}_1\, X_{1,i} + \tilde{b}_2\, X_{2,i} + \tilde{b}_4 \,X_{4,i})
\end{align}

Table \ref{tab:scen} shows the nine simulation scenarios that we assessed in our simulation study, which result from a fully factorial arrangement of the treatment effects on outcome and survival.
For simplicity, we assumed no drop-outs due to withdrawal of informed consent or loss to follow-up for other reasons than death.

\begin{table}[h]
  \caption{Overview of simulation scenarios.}\label{tab:scen}
  \begin{tabular}{l|rl|rl}
  \hline
    Scenario & \multicolumn{2}{c}{Treatment effect on Outcome} & \multicolumn{2}{|c}{Treatment effect on Survival}\\
             & \multicolumn{2}{c}{Mean difference (MD)}        & \multicolumn{2}{|c}{Odds ratio (OR)}\\
    \hline
    A        & 5 & (outcome increased) & 2 & (survival probability higher)\\
    B        & 5 & (outcome increased) & 1 & (no effect)\\
    C        & 5 & (outcome increased) & 0.5 & (survival probability lower)\\
    D        & 0 & (no effect) & 2 & (survival probability higher)\\
    E        & 0 & (no effect) & 1 & (no effect)\\
    F        & 0 & (no effect) & 0.5 & (survival probability lower)\\
    G        & -5 & (outcome decreased) & 2 & (survival probability higher)\\
    H        & -5 & (outcome decreased) & 1 & (no effect)\\
    I        & -5 & (outcome decreased) & 0.5 & (survival probability lower)\\
    \end{tabular}
  \end{table}

\subsection{Estimands}\label{sec:estimand}

The estimates derived by the three methods (see Section \ref{sec:methods.eval}) were compared in terms of bias, mean squared error and coverage with regard to two estimands.
The first estimand $\theta_1$ is the treatment effect on the outcome used in the simulation, equivalent to $b_z$ (see Section \ref{sec:m.dgm}), corresponding to the average causal effect $E[Y(1, \text{survival until 2 years}) - Y(0, \text{survival until 2 years})]$.
$\theta_1$ is a hypothetical estimand and represents the causal effect of the treatment on the outcome in the absence of mortality (up to 2 years of age). 
The second estimand $\theta_2$ is the survivor average causal effect (SACE), the treatment effect on the ``always survivors'', the patients who would have survived under both treatments, defined as $E[Y(1) - Y(0) \given S(1) = S(0) = 1]$.
$\theta_2$ is a principal stratum estimand and is relevant here, because mortality cannot be eliminated in the population at hand, and because a treatment effect on survival cannot be ruled out.
While $\theta_1$ is clearly a marginal estimand, $\theta_2$ may be seen as a marginal estimand for the subpopulation of ``always survivors'' or as a conditional expectation regarding survival status \citep{McGuinness2019, Luo2022}.
Note that the average treatment effect on survivors, which is estimated by complete case analysis, is defined as $E[Y(1) \given S(1) = 1] - E[Y(0) \given S(0) = 1]$.

While $\theta_1$ is defined by the simulation scenarios, $\theta_2$ is \emph{a priori} unknown.
We thus used the simulated observed and counterfactual data to derive $\theta_{2s}$ for each simulation $s=1,\dots, n_\text{sim}$.
Given the simulated observed outcome $Y_i(z)$ and the simulated counterfactual outcome $Y_i(1-z)$, the principal stratum of always survivors can be identified and $\theta_{2s}$ can be approximated as defined in Equation 1 of \cite{Hayden2005}:

\begin{equation}\label{eq:sace_ref}
\theta_{2s} = \frac{\frac1n\sum_i \{Y_i(1) - Y_i(0)\} S_i(1) S_i(0)}{\frac1n\sum_i  S_i(0) S_i(1)} = \frac{\sum_i \{Y_i(1) - Y_i(0)\} S_i(1) S_i(0)}{\sum_i  S_i(0) S_i(1)}.
\end{equation}

Here, $Y_i(1)$ and $Y_i(0)$ are the potential outcomes of patient $i$ on treatment and control, and $S_i(1)$ and $S_i(0)$ are the potential survival outcomes of patient $i$ under treatment and control, respectively (if $S_i(1) = 1$, patient $i$ survived on treatment, if $S_i(1) = 0$, patient $i$ died on treatment; if $S_i(0) = 1$ patient $i$ survived on control, if $S_i(0) = 0$ patient $i$ died on control).
To have one estimand per scenario, we then averaged over all simulations to derive $\theta_2$ as $\frac{1}{n_\text{sim}} \sum \limits_{s=1}^{n_\text{sim}} \theta_{2s}$. The resulting values of $\theta_2$ are shown in Figure \ref{fig:summary.measures}.
Note that $\theta_2$ is strictly speaking an estimate of the SACE but serves as the estimand in our simulation study. 
An alternative method to derive $\theta_2$ that was listed in the simulation study protocol was omitted, since it was less intuitive and results were very similar.

\subsection{Methods to be evaluated}\label{sec:methods.eval}

We used the SACE estimator proposed by \citet[][Equation 4]{Hayden2005}, hereafter referred to as SACE estimator, and compare it with complete case analysis and with analysis after multiple imputation.
Note that the SACE estimator in our notation, for a specific scenario $s$, is as follows:
\begin{equation}\label{eq:sace_est}
\hat{\theta}_{2s} = \frac{\sum_i Y_i(1) S_i(1) \hat{p}_i(0)}{\sum_i  S_i(1) \hat{p}_i(0)} - \frac{\sum_j Y_j(0) S_j(0) \hat{p}_j(1)}{\sum_j  S_j(0) \hat{p}_j(1)}
\end{equation}
where $i$ indexes over patients assigned to arm $z$ = 1 and $j$ indexes over patients assigned to arm $z$ = 0. 
The estimates $\hat{p}_i(0)$ and $\hat{p}_j(1)$ refer to the probability of survival under Placebo and Epo, as estimated by logistic regression from separate fits to the Placebo and Epo arms, respectively.
Complete case analysis was performed on the survivors only and is used as a benchmark for a biased analysis in most situations, since the treatment effect on survivors is not a randomized comparison. Complete case analysis was done using a linear model with treatment as explanatory variable.
Only in the absence of mortality or if mortality occurred ``completely at random'' would complete case analysis provide an unbiased estimator for both, $\theta_1$ and $\theta_2$, which then coincide with the treatment effect on survivors.
Given the underlying \emph{missing at random} assumption is reasonable and that multiple imputation can make use of available characteristics that are sufficiently associated with the outcome to be imputed, the treatment effect estimate derived by analysis of multiply imputed data can be expected to be close to $\theta_1$, independent of the treatment effect on survival.
Complete case analysis and the SACE estimator do not conform with the intention-to-treat (ITT) principle, as only a subset of patients is analyzed. 
In contrast, multiple imputation conforms with the ITT principle but creates unobservable (hypothetical) data.

As previously mentioned, the SACE is not identifiable without strong assumptions. A comprehensive overview of principal stratification methods and their underlying assumptions is given in \cite{Lipkovich2022}. 
\citeauthor{Hayden2005} make the \emph{stable unit treatment value} assumption \citep[SUTVA,][]{Rubin1980} and the \emph{explainable nonrandom survival} assumption, which they defined themselves and named it in the style of \emph{explainable nonrandom noncompliance} \citep{Robins1998}, despite some important differences \citep[see][]{Vansteelandt2024}.
The SUTVA implies that a subject's observed outcome is the same as (consistent with) the potential outcome associated with the treatment that the subject was assigned/randomized to and that the subject's potential outcomes do not depend on (or interfer with) the treatment assigned to other subjects \citep{Bornkamp2021, Lipkovich2022}. The SUTVA needs to be made by most principal stratification methods (and other causal inference approaches), since it allows to connect potential and observed outcomes at an individual patient level.
One reason for choosing Hayden's SACE estimator was that it neither relies on the \emph{monotonicity} assumption nor on the \emph{exclusion restriction} assumption (two other common assumptions), which are both unrealistic in the context of the SACE and the EpoRepair study. 
Monotonicity would imply that patients who die on the experimental treatment would also die on the control treatment, i.e., that there are no control-only survivors. 
\emph{Exclusion restriction} may be reasonable when the aim is to estimate the CACE, the average treatment effect on compliers, defined as those who would take the intervention when randomized to intervention and control when randomized to control both throughout the trial \citep[e.g., Table 1 in][]{Lipkovich2022}. Assuming \emph{exclusion restriction} in this setting would imply that potential outcomes for never-takers (those who never take the intervention) and always-takers (those who always take the intervention) are the same, regardless of the randomized treatment, as they actually take the same treatment.
In contrast, assuming \emph{exclusion restriction} for the SACE would imply that potential outcomes of always-survivors and never-survivors (sometimes called the doomed) would be the same.
This would be an unrealistic assumption, as it would imply no treatment effect in this stratum for which we want to estimate one.
Instead, Hayden's SACE estimator relies on modeling to identify strata membership based on observed baseline covariates, assuming \emph{explainable nonrandom survival}:
\begin{align}
\label{eq:nrs1}
S_i(z) &\independent S_i(1-z) \; \given \; X_i\\
\label{eq:nrs2}
S_i(z) &\independent Y_i(1-z) \; \given \; X_i, \{ S_i(1-z)=1 \}
\end{align}
The first assumption states that, conditional on the baseline covariates $X_i$, the survival status $S_i(z)$ of a subject under treatment $z$ is independent of its survival status $S_i(1-z)$ under treatment $1-z$.
The second assumption states that, conditional on surviving when assigned to treatment $1-z$, and on the baseline covariates $X_i$, the survival status $S_i(z)$ of a subject under treatment $z$ is independent of its outcome $Y_i(1-z)$ under treatment $1-z$.
However, since survival under treatment $z$ and $1-z$ can never be jointly observed, \emph{explainable nonrandom survival} may be called a ``cross-world'' assumption, which is not testable on the data.
In summary, \emph{explainable nonrandom survival} is a strong and untestable assumption \citep{Lipkovich2022, Kurland2009}.
\cite{Hayden2005} suggested and performed a sensitivity analysis for departures from assumptions (\ref{eq:nrs1}) and (\ref{eq:nrs2}), which they conducted on data from the ADRSnet clinical trial. They concluded that the significant treatment effect on days to return home (secondary outcome in ADRSnet, only observed in survivors) is fairly robust to departures from the independence assumptions, except in the presence of strong unappreciated interactions between covariates and treatment. 
While \cite{Qu2023} also concluded that assumptions (\ref{eq:nrs1}) and (\ref{eq:nrs2}) may be less implausible than monotonicity, these assumption are among those scrutinized by \cite{Vansteelandt2024}, who provide a thorough discussion of assumptions made in order to address principal stratum estimands, with a focus on treatment adherence (i.e., the CACE, since adherence and compliance are often used interchangeably).
However, in the case of the SACE, if we cannot assume exclusion restriction and do not want to assume monotonicity, we cannot avoid making the above or similar assumptions.
Further, in order to make the \emph{explainable nonrandom survival} assumption more plausible, one should collect and use baseline covariates which are strongly predictive for survival (which we tried), and ideally incorporate an analysis of sensitivity of results to departures from Equations (\ref{eq:nrs1}) and (\ref{eq:nrs2}).

In our simulation study we thus performed three different analyses using the SACE estimator. 
In the first analysis, the survival probabilities were estimated using all covariates that were used in the simulation of patient survival (gestational age, head circumference and Apgar score). This analysis ensures that the \emph{explainable nonrandom survival} assumption is met.
In the second analysis, we omitted the covariate head circumference and in the third analysis the covariate gestational age, which should lead to a violation of the \emph{explainable nonrandom survival} assumption. Since gestational age is more strongly related to survival than head circumference, the violation should be larger when omitting gestational age.

Multiple imputation of missing outcomes was performed using the R package \texttt{mice} \citep{vanBuuren2011}, generating 10 imputations per missing value.
Similar to the analysis using the SACE estimator, we once used all covariates that were used in the simulation of patient outcomes (gestational age, head circumference and SES) as predictors for the imputation of the missing outcome values, and omitted the covariate head circumference or gestational age in two additional analyses.
Here, the omission of these covariates should lead to a violation of the \emph{missing at random} assumption.
The analysis model was the same linear model used for complete case analysis, just applied to the imputed data.
Results were pooled across all imputations using Rubin's rules, as implemented in \texttt{mice}.

In addition to the methods described above, two alternative methods to estimate the SACE, or more precisely, upper and lower bounds of the SACE \citep{Zhang2003, Chiba2011}, are described and applied to our simulations in the supporting information. These methods rely on the \emph{monotonicity} assumption.

\subsection{Performance measures}\label{sec:m.perf}

For each simulation $s=1,\dots, n_\text{sim}$, the treatment effect on the outcome was estimated with each method $m$.
The resulting treatment effect $\hat{\theta}_{sm}$ was stored together with its standard error.
Based on these numbers, 95\,\% Wald confidence intervals $[L_{sm}, U_{sm}]$ were calculated \citep[as recommended by][]{Burton2006}.
Then, within each of the 9 scenarios, we calculated the average of the treatment effect estimates for each method based on $\hat{\theta}_{sm}$, as 
\begin{equation}
\bar{\hat{{\theta}}}_{m}=\frac{1}{n_\text{sim}}\sum\limits_{s=1}^{n_\text{sim}}\hat{\theta}_{sm}, 
\end{equation}
together with a Wald confidence interval based on the empirical standard error of $\bar{\hat{{\theta}}}_{m}$, as well as several performance measures.
An index for the scenario is omitted for simplicity.

For each scenario, the performance of each statistical method $m$, was evaluated in terms of bias, mean squared error (MSE) and coverage with respect to the estimand $\theta_k$, $k=1,2$. 
Each performance measure was calculated with a Monte Carlo standard error as shown in Table \ref{tab:perf}, following the definitions given in \cite{Morris2019}. 
The bias is a systematic difference between the result of a specific method (estimator) from the estimand.
We calculated the observed bias for each scenario and method with regard to each estimand as the average difference between the estimates and the estimand.
The MSE is a measure of the accuracy of a method, which can also be written as the sum of the variance of the estimator and the squared bias of the estimator. 
This implies that in the case of unbiased estimators, the MSE and variance are equivalent.
The MSE accounts for the fact that there is typically a trade-off between bias and variance of a method.
The coverage is the proportion how often the 95\,\% confidence interval for the treatment effect estimate $[L_{sm}, U_{sm}]$, contained the estimand over all 1300 simulations per method.

The number of simulations, $n_\text{sim}$, to perform for each scenario was calculated based on the accuracy of the SACE estimate, using 

\begin{equation*}
n_\text{sim}=\bigg( \frac{Z_{1-\alpha/2} \cdot \sigma}{\delta} \bigg)^2 
\end{equation*}

where $\delta$ is the specified level of accuracy of the SACE estimate we were willing to accept, i.e. the permissible difference from the true value, $Z_{1-\alpha/2}$ is the quantile of the standard normal distribution and $\sigma$ is the standard error of the SACE estimate, which can be obtained from the real data \citep{Burton2006}.
From the EpoRepair trial we estimated $\sigma=4.4$ using 109 patients for whom the outcome was available when alive and for whom all covariates needed to estimate the SACE were available, including an estimated number of 77 always survivors.
Because we simulated data for $n=500$ patients per simulation, we expected a smaller standard error by a factor of $\sqrt{\frac{77}{500}}$, resulting in $\sigma=1.73$.
To simulate data with a treatment effect of 5 (or 0 or -5) and to achieve an accuracy $\delta$ of 2\% ($5 \cdot 0.02 = 0.1$) at a significance level $\alpha$ of 5\,\%, at least 1150 simulations would have been required.
We decided to round up and performed $n_\text{sim}=1300$ simulations.

\begin{table}[h]
  \caption{Formulas for the performance measures bias, mean squared error (MSE) and coverage of method $m$ with respect to estimand $\theta_k$: estimates and Monte Carlo standard errors (SE).}\label{tab:perf}
  \centering
  \begin{footnotesize}
  \begin{tabular}{lll}
  \hline
    Measure & Estimate & Monte Carlo SE\\
    \hline
    Bias & $\frac{1}{n_\text{sim}}\sum\limits_{s=1}^{n_\text{sim}}\hat{\theta}_{sm} - \theta_k$ & $\sqrt{\frac{1}{n_\text{sim}(n_\text{sim}-1)}              \sum\limits_{s=1}^{n_\text{sim}}(\hat{\theta}_{sm} - \bar{\hat{{\theta}}}_m)^2}$\\
    MSE & $\frac{1}{n_\text{sim}} \sum\limits_{s=1}^{n_\text{sim}}(\hat{\theta}_{sm} - \theta_k)^2$ & 
    $\sqrt{\frac{\sum\limits_{s=1}^{n_\text{sim}}\big[ (\hat{\theta}_{sm} - \theta_k)^2-\widehat{\text{MSE}} \big]^2}{n_\text{sim}(n_\text{sim}-1)}}$ \\
    Coverage & $\frac{1}{n_\text{sim}}\sum\limits_{s=1}^{n_\text{sim}} \mathbb{1}\{L_{sm} \leq \theta_k \leq U_{sm}\}$ & $\sqrt{\frac{\widehat{\text{Coverage}}(1- \widehat{\text{Coverage}})}{n_\text{sim}}}$ \\
    \end{tabular}
  \end{footnotesize}
  \end{table}

\section{Results}

\subsection{Number of patients analyzed}
The average number of patients analyzed per scenario mainly depended on the method of analysis and on the treatment effect on survival (Table \ref{tab:n.analyzed}). 
Across scenarios with the same treatment effect on survival, the average number of patients analyzed was highest for multiple imputation (all patients, $n=500$), intermediate for complete case analysis (all survivors) and lowest for the SACE estimator (principal stratum of always survivors).
Regarding the SACE, it should be noted that the average number of patients analyzed as reported in Table \ref{tab:n.analyzed} is rather the average effective sample size on which the SACE estimates are based (average estimated number of always survivors) which corresponds to the sum of the two denominators in Equation \ref{eq:sace_est}. Whereas the survival probabilities are estimated for all $n=500$ patients, the estimated number of always survivors is reduced through conditioning on observed survival (e.g., $S_i(1)$), and weighting by the survival probability on the counterfactual treatment (e.g., $\hat{p}_i(0)$).
The estimated size of the principal stratum of always survivors was very similar when the SACE was estimated using all covariates or when head circumference was omitted, but was slightly smaller when gestational age was omitted.
The percentage of survivors (or the percentage mortality) and always survivors depended on the treatment effect on survival (but not on the treatment effect on outcome) and was lowest in scenarios with a positive treatment effect on survival (OR=2, scenarios A, D, and G), intermediate without an effect on survival (OR=1, scenarios B, E and H) and highest in scenarios with negative effect on survival (OR=0.5, scenarios C, F and I). 
These differences in the number of patients analyzed (per simulation) affected the average treatment effect estimates and thus the performance measures, but hardly affected the confidence intervals of these quantities, which mostly depend on the variability between simulations per scenario ($n_\text{sim}=1300$).

\begin{table}[ht]
\centering
\caption{Average number of patients analyzed per simulated trial depending on the simulated treatment effect on survival (TE on S, rows) and the method of analysis (columns 1--5), as well as the percentage survivors (column 6) and always survivors (AS, columns 7--9), depending on the treatment effect on survival. For the SACE estimator the average number of patients analyzed as well as the percentage always survivors is shown for the analysis using all covariates, the analysis omitting head circumference (no hc) or gestational age (no ga). 
} 
\label{tab:n.analyzed}
\begingroup\footnotesize
\begin{tabular}{lrrrrrrrrr}
  \hline
TE on S & MI ($n$) & CCA ($n$) & SACE ($n$) & SACE no hc ($n$) & SACE no ga ($n$) & Survivors (\%) & AS (\%) & AS no hc (\%) & AS no ga (\%) \\ 
  \hline
OR=2 & 500 & 432.4 & 389.3 & 389.0 & 382.2 & 86.5 & 77.9 & 77.8 & 76.4 \\ 
  OR=0 & 500 & 418.3 & 370.3 & 369.9 & 361.3 & 83.7 & 74.1 & 74.0 & 72.3 \\ 
  OR=0.5 & 500 & 399.5 & 342.8 & 342.4 & 332.5 & 79.9 & 68.6 & 68.5 & 66.5 \\ 
   \hline
\end{tabular}
\endgroup
\end{table}

\subsection{Treatment effect estimates and bias}\label{sec:r.te.bias} 

\begin{figure}[h]
\includegraphics[width=\textwidth]{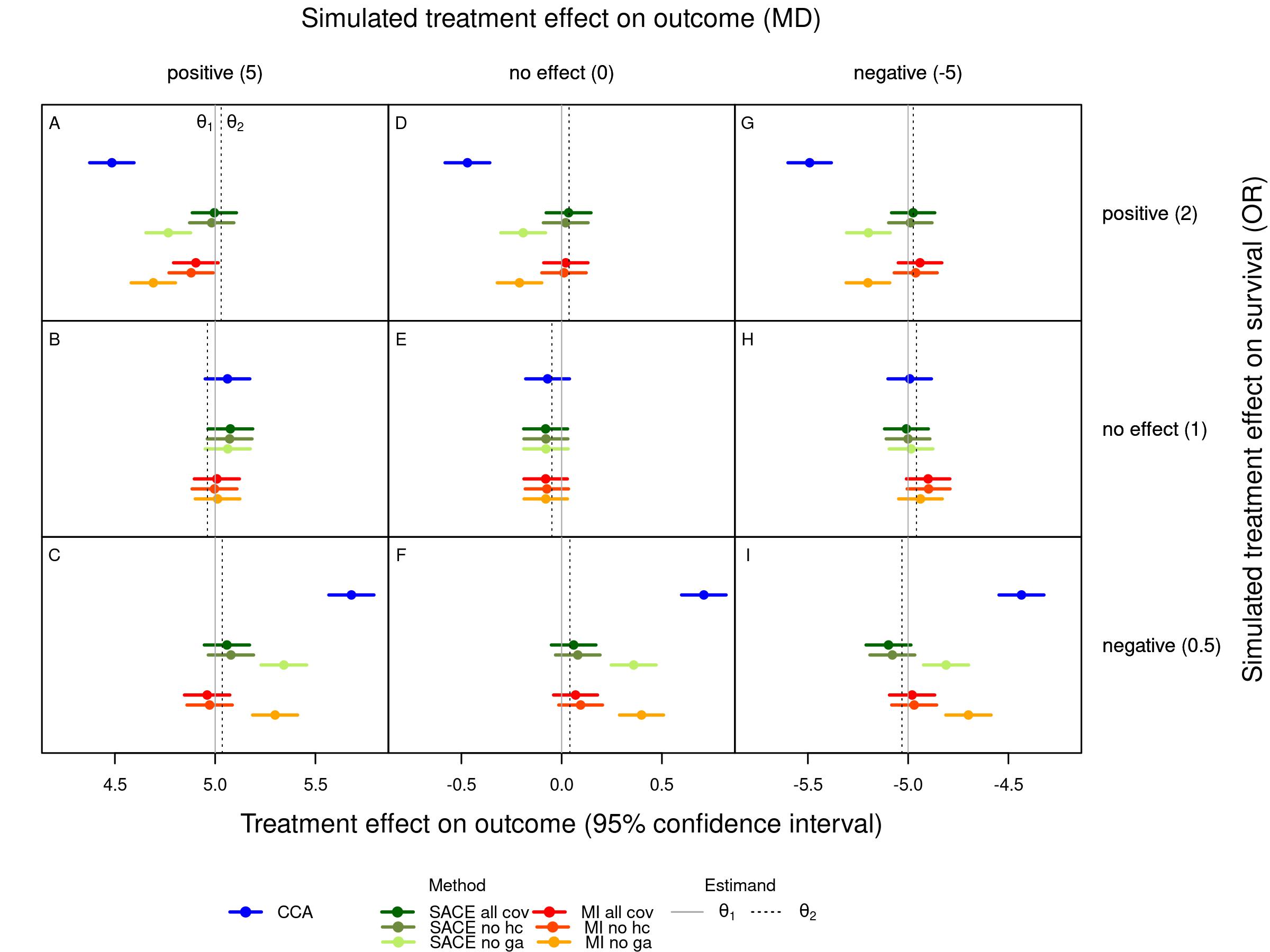}
\caption{Average treatment effect estimates with 95\,\% Wald confidence intervals for each scenario (A--I) and method. For the SACE estimator and for multiple imputation, the average estimates from the analyses using all covariates (green and red) are shown together with those from the analyses where either the covariate head circumference or gestational age was omitted (lighter shades of green and red). The scenarios are arranged with the treatment effect on survival in rows and the treatment effect on the outcome in columns. The vertical lines indicate the estimands, i.e., the treatment effect on the outcome used in the simulation, $\theta_1$ (solid gray line), and the SACE derived from observed and counterfactual data, $\theta_2$ (dashed black line).
Note the different x-axis scales for the three columns (with different simulated treatment effects on outcome).}\label{fig:summary.measures}
\end{figure}

Figure \ref{fig:summary.measures} shows the average estimates of the treatment effect on the outcome for each scenario and method together with the estimands $\theta_1$ and $\theta_2$.
For the SACE estimator and for multiple imputation, the average estimates from the analysis using all covariates (green and red) are shown together with those from the analyses omitting either of the covariates head circumference or gestational age (two lighter shades of green and red).
Across all scenarios the two estimands, the causal effect of the treatment in the absence of mortality ($\theta_1$) and the survivor average causal effect ($\theta_2$), were similar.
In scenarios without a treatment effect on survival (B, E, H, middle row of Figure \ref{fig:summary.measures}), treatment effect estimates were similar for all methods and close to the causal effect used in the simulation.
In particular, in the absence of a treatment effect on both outcome and survival (E) all methods estimated very similar treatment effects, even in presence of a significant proportion of outcomes truncated by death. Since all estimands are equivalent under these conditions, this result could be expected.

In scenarios where Epo improved survival compared to placebo (A, D, G, top row of Figure \ref{fig:summary.measures}), complete case analysis consistently estimated smaller or more negative treatment effects than the SACE estimator and multiple imputation, and as a consequence, also small negative effects in case of no causal treatment effect on the outcome.
Thus, complete case analysis underestimated the positive treatment effect on the outcome (A), overestimated the negative treatment effect (G) and estimated a small negative effect when there was actually no effect (D).
These results may be explained by the better survival of frail patients, e.g., preterm infants with lower gestational age, also called ``frail mortality benefiters'' by \cite{Colantuoni2018}, under Epo compared to placebo.
When Epo reduced survival compared to placebo (C, F, I, bottom row of Figure \ref{fig:summary.measures}), complete case analysis estimated larger or less negative treatment effects than the other two methods, and even small positive effects in case of no causal treatment effect on the outcome.
Thus, complete case analysis overestimated the positive treatment effect on the outcome (C), underestimated the negative treatment effect (I) and estimated a small positive effect when there was actually no effect (F). 
These results may be explained by the increased mortality of frail patients under Epo compared to placebo, resulting in better average outcome in the Epo group.
When head circumference was omitted, the SACE estimator and multiple imputation resulted in very similar treatment effect estimates as when all covariates were used.
However, when gestational age was omitted, estimates were shifted towards those from complete case analysis.

Figure 1 in the supporting information is analogous to Figure \ref{fig:summary.measures}, but additionally shows average bounds for the SACE estimated using the method of \cite{Chiba2011} and the method of \cite{Zhang2003}.
Table 3 in the supporting information summarizes for each scenario how many times (out of the total of 1300 the SACE estimator \citep[based on][]{Hayden2005} lay between the bounds based on the two alternative methods.
In scenarios where the \emph{monotonicity} assumption made sense (all except, B, E and H), the proportion of SACE estimates
that lay in the expected range was between 84 and 90\,\%.
Some deviations from the \emph{monotonicity} assumption occurred in all scenarios with a treatment effect on
survival, since the strata assumed to be empty were never completely empty (Table 4 in the supporting information).

\begin{figure}
\includegraphics[width=0.9\textwidth]{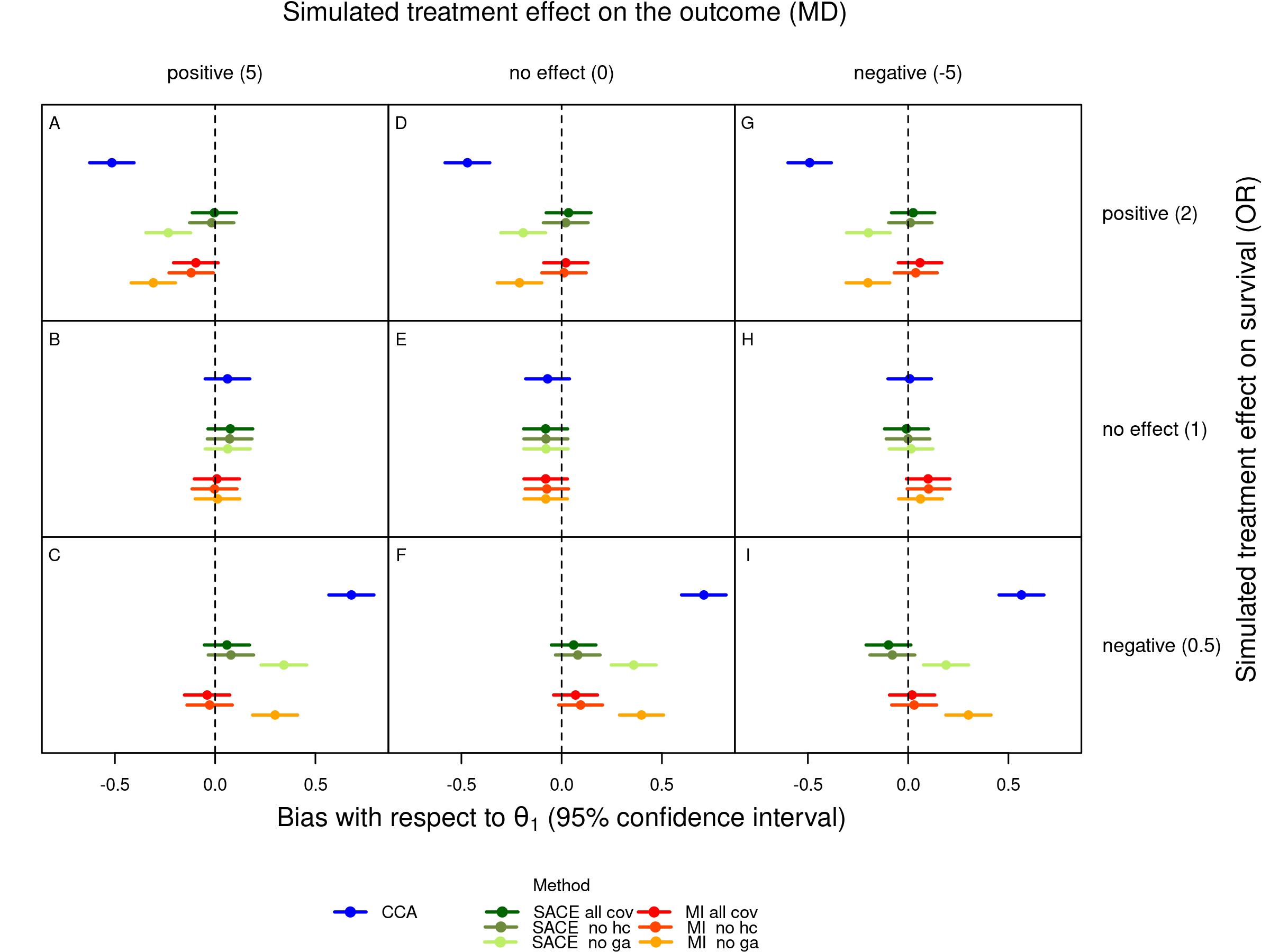}
\includegraphics[width=0.9\textwidth]{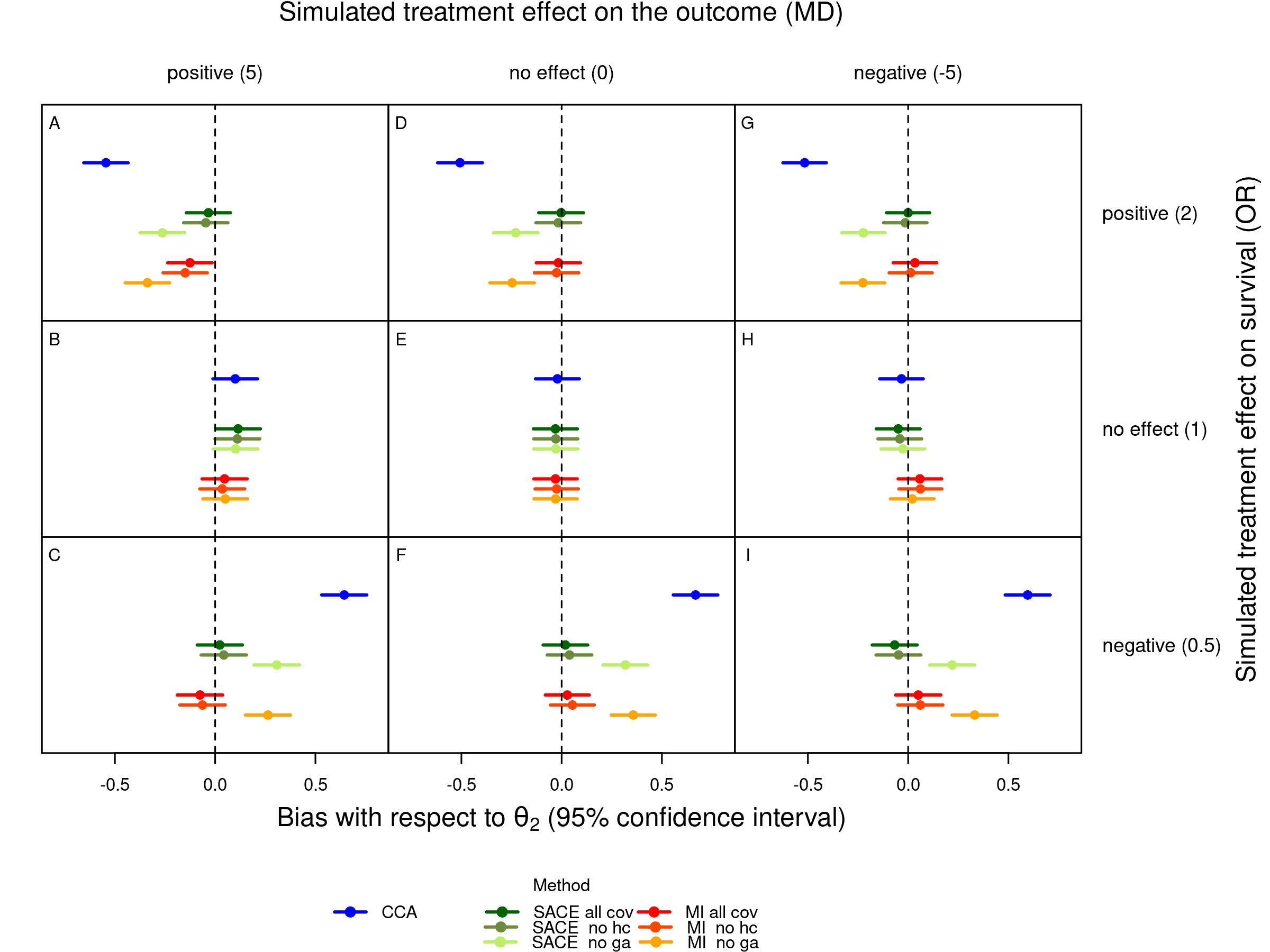}
\caption{Average bias for each scenario (A--I) and method with regard to the treatment effect on the outcome used in the simulation, $\theta_1$ (top panels) and with regard to the SACE derived from observed and counterfactual data, $\theta_2$ (bottom panels). Bias is shown with 95\,\% Monte Carlo confidence interval. For the SACE estimator and for multiple imputation, the bias of the analyses using all covariates (green and red) is shown together with the bias of the analyses where either the covariate head circumference or gestational age was omitted (lighter shades of green and red). The scenarios are arranged with the treatment effect on survival in rows and the treatment effect on the outcome in columns. The dashed vertical lines indicate no bias.}\label{fig:bias.theta1vs2}
\end{figure}

The top part of Figure \ref{fig:bias.theta1vs2} shows the average bias with 95\,\% Monte Carlo confidence interval for each scenario and method with regard to $\theta_1$, the treatment effect on the outcome used in the simulation. 
Likewise, the bottom part of Figure \ref{fig:bias.theta1vs2} shows the average bias with regard to  $\theta_2$.
Note that the patterns are almost the same as those shown in Figure \ref{fig:summary.measures}, just on the scale of the bias instead of the outcome.

When survival was affected by treatment, there was a marked average bias from complete case analysis with regard to $\theta_1$ and $\theta_2$, with 95\% confidence intervals that clearly exclude a bias of $0$. 
The bias was negative when Epo improved survival compared to placebo, and positive when Epo reduced survival compared to placebo.
Further, the bias of complete case analysis with regard to $\theta_1$, was largest in the scenarios with the highest mortality (lowest percentage of survivors and always survivors in Table \ref{tab:n.analyzed}) and thus the smallest number of patients analyzed, and accounted for approx. 10\,\% and 13\,\% of $\theta_1$ in these scenarios (bottom rows of the top and bottom parts of Figure \ref{fig:bias.theta1vs2}, Table \ref{tab:av.bias}).
When survival was not affected by treatment, the 95\,\% confidence intervals for the average bias included 0.

As already indicated in Figure \ref{fig:summary.measures}, the average bias of the SACE estimator and multiple imputation with regard to both $\theta_1$ and $\theta_2$ was much lower, compared to the bias of complete case analysis, when all covariates were used or when head circumference was omitted.
Omitting gestational age considerably increased the bias of the SACE estimator and multiple imputation, with 95\,\% confidence intervals that exclude a bias of $0$, but the bias of these analyses was still much smaller than that of complete case analysis.
Since the SACE estimator targets $\theta_2$ (estimated as $\hat{\theta}_{2}$), and multiple imputation targets $\theta_1$, we expected that the corresponding bias would be lower for these combinations of method and estimand than for the others.
However, although the average bias of the SACE estimator with regard to $\theta_2$ and of multiple imputation with regard to $\theta_1$ was small in all scenarios (Table \ref{tab:av.bias}), and 95\,\% confidence intervals for bias included 0, except for analyses omitting gestational age (Figure \ref{fig:bias.theta1vs2}), the expected patterns were not consistently observed.
The summary Table \ref{tab:av.bias} indicates that multiple imputation was not generally less biased with regard to $\theta_1$ than $\theta_2$. 
Particularly in scenarios with a negative treatment effect on survival (rows with OR=0.5 in Table \ref{tab:av.bias}), this pattern was sometimes even reversed.
This may be due to the larger mortality in these scenarios which resulted in smaller numbers of patients analyzed by the SACE estimator and larger number of missing values which were multiply imputed.
With regard to $\theta_2$, multiple imputation was often more biased than the SACE estimator.
Further, the analyses omitting the covariate head circumference or gestational age with the SACE estimator resulted in slightly larger and much larger average bias with regard to $\theta_2$ than the corresponding analysis with all covariates, and the same applied for multiple imputation with regard to $\theta_1$.
The observed increase in bias could be expected due to violation of the \emph{explainable nonrandom survival} assumption relevant for $\theta_2$ and the \emph{missing at random} assumption relevant for $\theta_1$.

\begin{table}[ht]
\centering
\caption{Average bias regarding $\theta_1$ and $\theta_2$ depending on the odds ratio (OR) for the treatment effect on survival (rows) and the method of analysis (columns 3--9).} 
\label{tab:av.bias}
\begin{tabular}{ll|r|rrr|rrr}
  \hline
Estimand & OR & CCA & SACE all cov & SACE no hc & SACE no ga & MI all cov & MI no hc & MI no ga \\ 
  \hline
$\theta_1$ & 2 & $-$0.4920 &  0.0185 &  0.0047 & $-$0.2079 & $-$0.0053 & $-$0.0234 & $-$0.2396 \\ 
  $\theta_1$ & 0 & $-$0.0003 & $-$0.0042 & $-$0.0023 & $-$0.0003 &  0.0094 &  0.0084 & $-$0.0020 \\ 
  $\theta_1$ & 0.5 &  0.6512 &  0.0069 &  0.0268 &  0.2970 &  0.0163 &  0.0324 &  0.3326 \\ 
   \hline
$\hat{\theta}_2$ & 2 & $-$0.5226 & $-$0.0121 & $-$0.0259 & $-$0.2385 & $-$0.0359 & $-$0.0540 & $-$0.2702 \\ 
  $\hat{\theta}_2$ & 0 &  0.0153 &  0.0113 &  0.0133 &  0.0153 &  0.0250 &  0.0239 &  0.0135 \\ 
  $\hat{\theta}_2$ & 0.5 &  0.6359 & $-$0.0084 &  0.0115 &  0.2817 &  0.0010 &  0.0171 &  0.3173 \\ 
   \hline
\end{tabular}
\end{table}

\subsection{Mean squared error}

\begin{figure}
\includegraphics[width=0.9\textwidth]{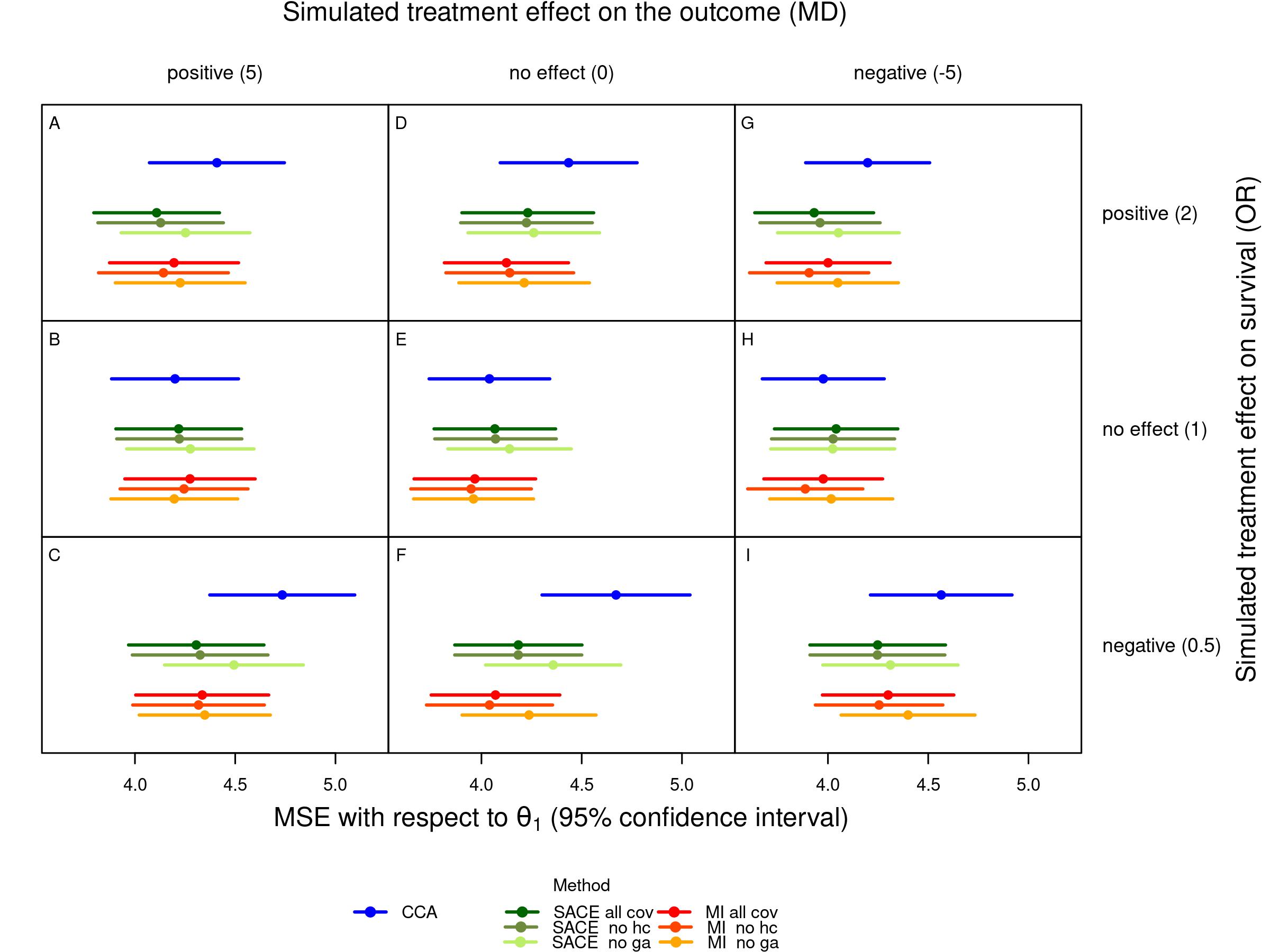}
\includegraphics[width=0.9\textwidth]{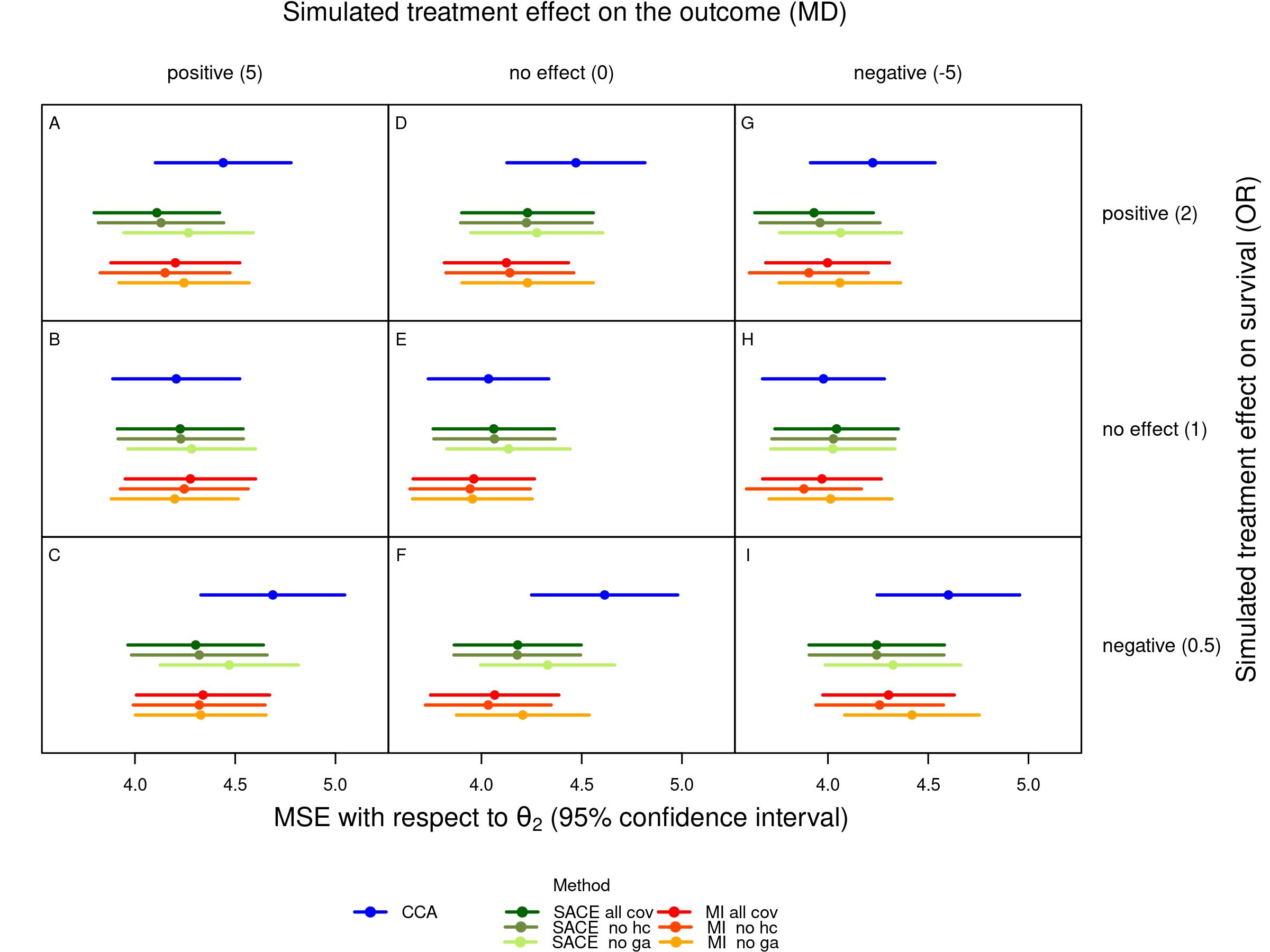}
\caption{Average mean squared error (MSE) for each scenario (A--I) and method with regard to the treatment effect on the outcome used in the simulation, $\theta_1$ (top panels) and with regard to the SACE derived from observed and counterfactual data, $\theta_2$ (bottom panels). MSE is shown with 95\,\% Monte Carlo confidence interval. For the SACE estimator and for multiple imputation, the MSE the analyses using all covariates (green and red) is shown together with the MSE of the analyses where either the covariate head circumference or gestational age was omitted (lighter shades of green and red). The scenarios are arranged with the treatment effect on survival in rows and the treatment effect on the outcome in columns.}\label{fig:mse.theta1vs2}
\end{figure}

The top part of Figure \ref{fig:mse.theta1vs2} shows the average MSE for each scenario and method with regard to $\theta_1$, with the 95\,\% Monte Carlo confidence interval. 
Likewise, the bottom part of Figure \ref{fig:mse.theta1vs2} shows the average MSE with regard to $\theta_2$.
The average MSE only weakly depended on the treatment effect on the outcome, which can be seen from the very similar patterns in all three columns of Figure \ref{fig:mse.theta1vs2}, but it strongly depended on the treatment effect on survival and on the method of analysis (rows and colors in Figure \ref{fig:mse.theta1vs2}).
As a consequence of the bias which was always largest for complete case analysis, this method also had the largest average MSE of all methods in the presence of treatment effects on survival.
The average MSE was similar and relatively small for all methods (and types of analyses) in scenarios without a treatment effect on survival (B, E, H), and largest in scenarios where Epo reduced survival compared to placebo (C, F, I, bottom rows of both parts of Figure \ref{fig:mse.theta1vs2}).
The latter is again a consequence of the larger mortality in these scenarios (Table \ref{tab:n.analyzed}), which also resulted in larger bias with regard to $\theta_1$ and $\theta_2$.
As for bias, the average MSE of analyses using all covariates and those omitting head circumference was very similar, whereas the MSE of analyses omitting gestational age was increased, but still smaller than the MSE of complete case analysis.

\subsection{Coverage}

\begin{figure}
\includegraphics[width=0.9\textwidth]{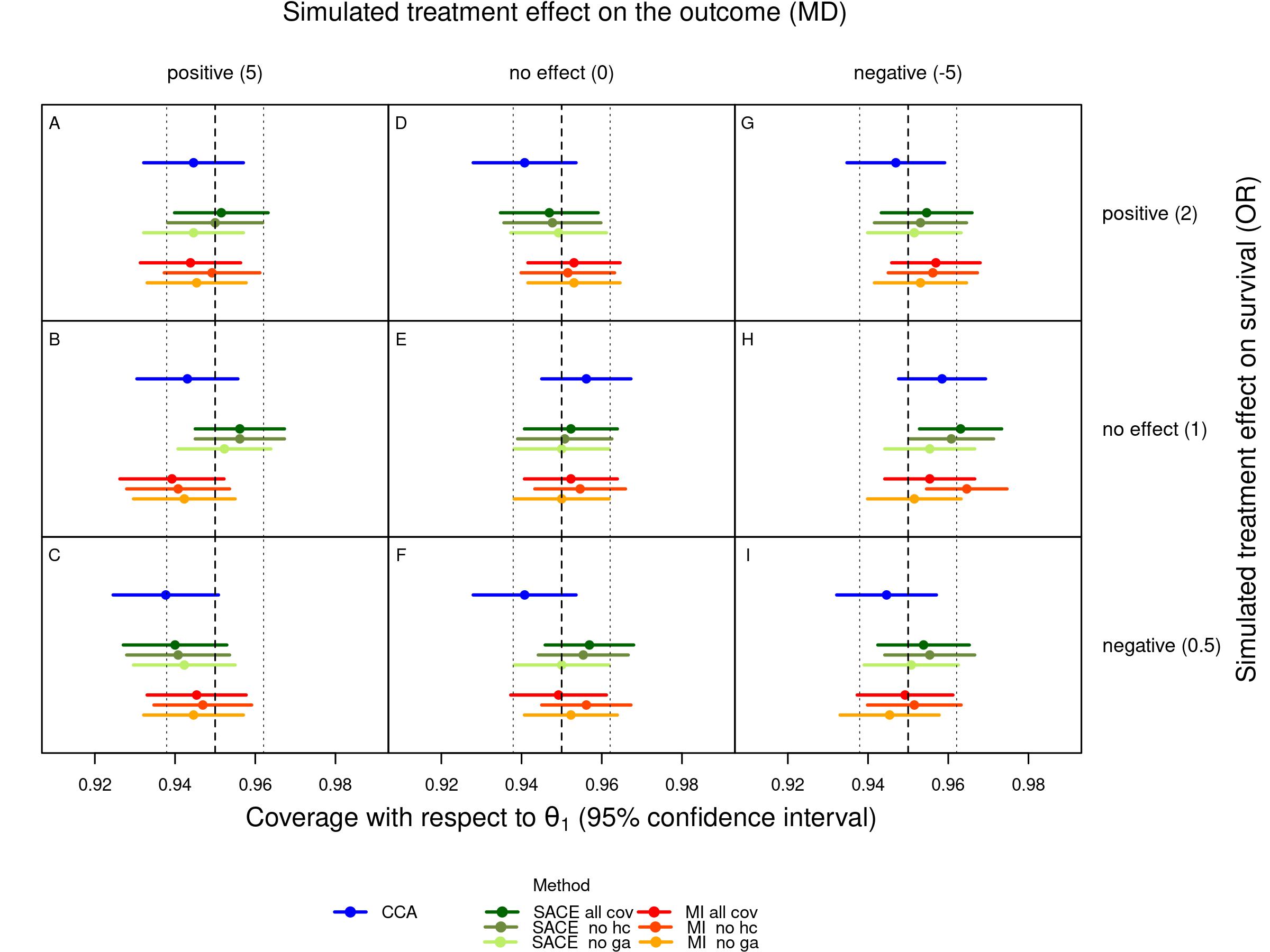}
\includegraphics[width=0.9\textwidth]{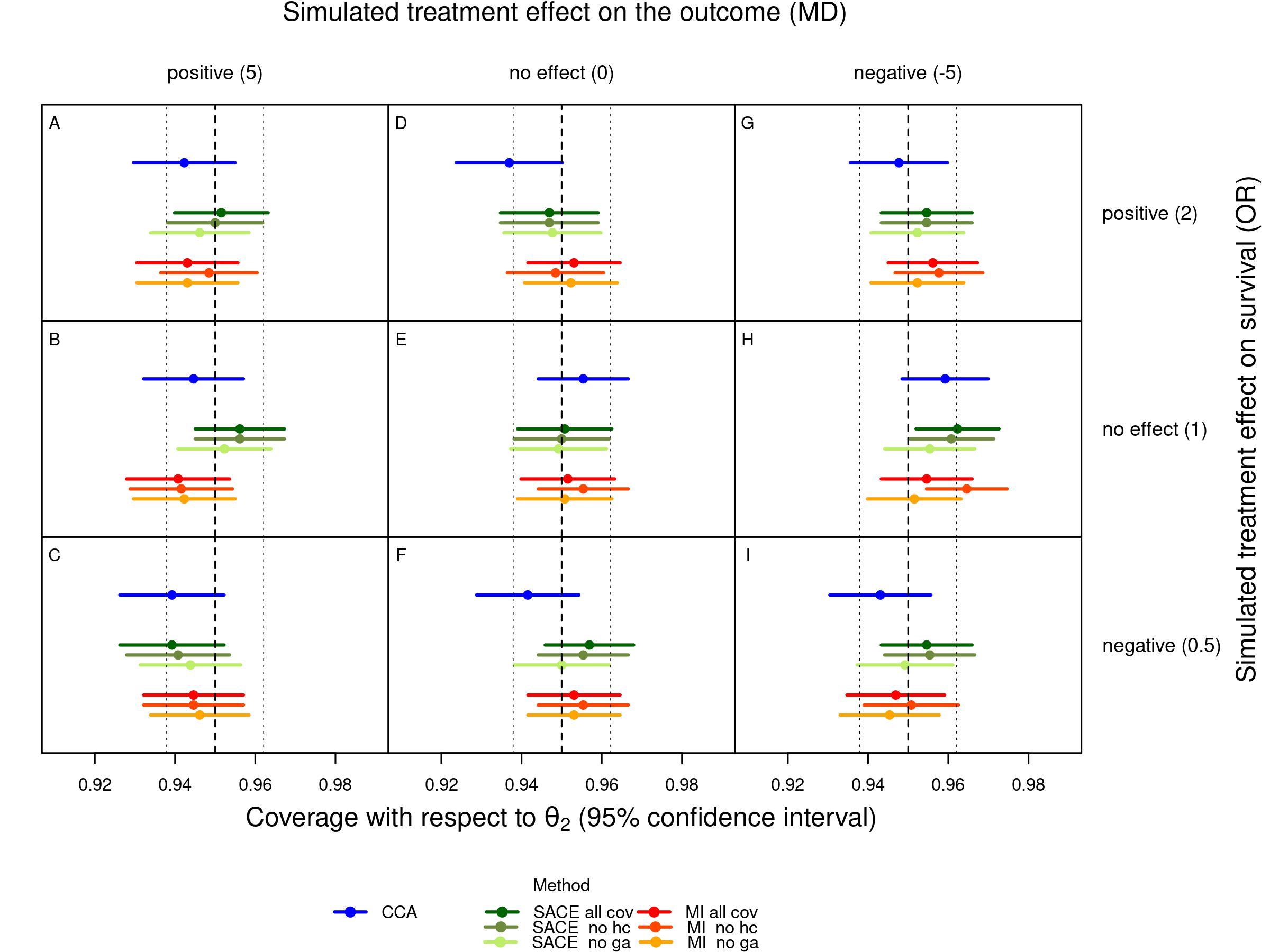}
\caption{Coverage for each scenario (A--I) and method with regard to the treatment effect on the outcome used in the simulation, $\theta_1$ (top panels) and with regard to the SACE derived from observed and counterfactual data, $\theta_2$ (bottom panels). Coverage is shown with 95\,\% Monte Carlo confidence interval. For the SACE estimator and for multiple imputation, the coverage of the analyses using all covariates (green and red) is shown together with the coverage of the analyses where either the covariate head circumference or gestational age was omitted (lighter shades of green and red). The scenarios are arranged with the treatment effect on survival in rows and the treatment effect on the outcome in columns. The dashed vertical lines indicate a coverage of 0.95, the dotted vertical lines indicate the acceptable coverage range (0.938 to 0.962).}\label{fig:cov.theta1vs2}
\end{figure}

The top part of Figure \ref{fig:cov.theta1vs2} shows the coverage for each scenario and method with regard to $\theta_1$, with the 95\,\% Monte Carlo confidence interval. 
Likewise, the bottom part of Figure \ref{fig:cov.theta1vs2} shows the coverage with regard to $\theta_2$.
\cite{Tang2005} suggested that coverage can be expected to lie within the interval of the nominal coverage probability $\pm$ standard error, which is the standard error of a proportion, i.e., $\sqrt{\frac{p(1-p)}{n}}$, here $\sqrt{\frac{0.95 \cdot 0.05}{1300}}$, and that values below this interval can be interpreted as under-coverage and values above as over-coverage.
In the context of our simulation study, under- and over-coverage mean that the corresponding estimand is captured less and more often, respectively.
Accordingly, the acceptable coverage range is 0.938 to 0.962 in our case, which is shown in Figure \ref{fig:cov.theta1vs2} (dotted vertical lines).
Similarly, we can check whether the nominal confidence-level of 0.95 is included in the observed 95\,\% Monte Carlo confidence intervals for coverage.
Both conditions are met with respect to both estimands in most cases.
Exceptions are (1) complete case analysis with regard to $\theta_1$ in scenario C and with regard to  $\theta_2$ in scenario D, where the point estimate for coverage lies below the acceptable range and the 95\,\% CI for coverage lies almost entirely below 0.95, indicating under-coverage and (2) the SACE estimator using all covariates in scenario H with regard to $\theta_1$ and $\theta_2$, where the point estimate for coverage lies above the acceptable range and the 95\,\% CI for coverage lies entirely above 0.95, indicating over-coverage, and (3) for multiple imputation omitting head circumference in scenario H, similar to (2).
Again as a consequence of the bias of complete case analysis with regard to $\theta_1$ and $\theta_2$, this method has a lower coverage than the other methods in the presence of treatment effects on survival.
The coverage of the analyses with all covariates and those either omitting head circumference or gestational age did not show a very clear pattern, for example, coverage of analyses omitting gestational age was not always lower than coverage of the other analyses, except for scenarios with a negative treatment effect on the outcome.

\clearpage

\section{Discussion}

We investigated the setting of an RCT with a continuous outcome truncated by death and performed a simulation study to compare treatment effect estimates from complete case analysis, analysis after multiple imputation of the missing outcome values and the SACE estimator proposed by \cite{Hayden2005}.
Both alternative approaches to complete case analysis, although targeting different estimands (principal stratum estimand, hypothetical estimand), efficiently reduced the bias compared to complete case analysis and led to similar estimates of the treatment effect. 
While covariate information was used to reduce the analysis data set to the principal stratum of ``always survivors'' by the SACE estimator, it was was also used to substitute the outcome measurements truncated by death by multiple imputation.
Further, we showed that results of the SACE estimator and multiple imputation were robust to minor violations of the ``explainable nonrandom survival'' or the \emph{missing at random} assumption due to model misspecification (i.e., omission of the covariate head circumference at birth in the analysis although it was used to generate the data), but were less robust to stronger violations (i.e., omission of gestational age). 

Our results confirm and illustrate that complete case analysis is not a favorable option when outcomes are truncated by death, unless a treatment effect on survival can be virtually ruled out. 
However, the question remains whether the SACE as a principal stratum estimand or a hypothetical estimand (ignoring death) are more meaningful in this situation, or whether even both approaches have their justification.
The relevance of the hypothetical estimand, the causal effect of the treatment on the outcome in the absence of mortality, is questioned because death could not be avoided by design in a future trial on preterm infants.
Conversely, principal stratum estimands are controversial. 
While they have been advocated by \cite{Frangakis2002}, have been viewed as the only sensible estimand when outcomes are truncated by death \citep{Rubin2006, VanderWeele2011}, and have seemingly taken a role in drug development \citep{Akacha2017, Qu2021a, Bornkamp2021}, they have been strongly criticised \citep[][see Introduction]{Robins2007, Pearl2011, Joffe2011, Dawid2012, Stensrud2023}.
%
This controversy with regard to principal stratum estimands may be seen as an argument in favour of a hypothetical estimand.
However, because both estimands have their clear shortcomings, we find it hard to clearly prefer one above the other. 
In situations when both may be used, we recommend to include both in the statistical analysis plan and to report and discuss the differences in interpretation.
Whenever the survivor average causal effect is used as an estimand, mortality should also be considered as an outcome. It is important to report the mortality overall and whether it differs between treatments, in line with recommendations by \cite{Akacha2017} to address adherence together with the treatment effect in adherers.

However, there are of course further options than hypothetical and principal stratum estimands in RCTs with outcomes truncated by death.
\cite{Stensrud2023} suggested ``conditional separable effects'' as alternative estimands to unravel treatment effects on a posttreatment variable such as death and treatment effects on an outcome of interest.
This is an elegant alternative to principal stratum estimands, in particular for oncology trials using chemotherapy with two components, as in the example given in their article.
It will be very interesting to see how such alternatives are taken up by trialists and biostatisticians in pharmaceutical and academic clinical research.
However, for the relatively small EpoRepair trial with an intervention consisting of one component only, a limited number of deaths and without repeated outcome measurements, application of such a complex method may overreach.
A further, much simpler alternative that is prevalent in neonatology is to use a composite endpoint approach, combining an unfavourable event with death.
Such unfavourable events are often binary, such as necrotizing enterocolitis or retinopathy of prematurity, but cognitive development is sometimes also dichotomized in cognitive impairment (score below a certain threshold) vs. normal cognitive development.
We could thus define a composite outcome as the occurrence of cognitive impairment or death (up to two years) and target a treatment strategy estimand, since all randomized patients can then be included in the analysis.
However, dichotomization of cognitive development leads to a loss in power.
A further disadvantage of this approach is that it combines "events" that are quite different, which is why \cite{Engel2016} suggested to separately consider healthy outcome, event of interest, and death.

Regarding the observability of principal strata, an interesting approach was used by \cite{Qu2023}. 
They argued that the potential outcomes of patients under both treatments can be observed in a cross-over trial when carry-over and period effects can be excluded.
Hence, they used a 2$\times$2 cross-over trial to evaluate commonly used assumptions to estimate principal stratum effects, including \emph{monotonicity}, within-treatment principal ignorability, and cross-world assumptions of principal ignorability and principal strata independence.
In line with our results in Table 4 in the supporting information, they found that the \emph{monotonicity} assumption did not hold well for two principal stratum variables. 
Moreover, they found that ``cross-world principal ignorability'' and ``cross-world principal stratum independence conditional on baseline covariates'', which they defined as two components of ``explainable nonrandom survival'' \citep[as required by][]{Hayden2005}, seemed reasonable. 
However, this is evidence from a single study only, and it is contrasted by a recent theoretical assessment of these assumptions by \cite{Vansteelandt2024}, who judge them as implausible. In particular, assumption (\ref{eq:nrs1}) is shown to be invalid under the null hypothesis of no treatment effect.

There are many other methods than the one suggested by \cite{Hayden2005} to estimate the SACE.
We chose this method because it neither assumes \emph{monotonicity} nor \emph{exclusion restriction}, which we both found unrealistic for the EpoRepair trial. 
In fact, in the context of the SACE in an RCT, \emph{monotonicity} may always be questionable because if one treatment is believed to benefit survival a priori, a clinical trial would be unethical \citep{WangL2017}.
However, avoiding \emph{monotonicity} and \emph{exclusion restriction} comes at the cost of making the ``explainable nonrandom survival'' assumption.
As other cross-world assumptions, it is untestable using observed data, as it involves both potential outcomes of patients.
Due to the simulation of both potential outcomes for each patient and subsequent ideal analysis using all covariates used in the simulation to model the survival probabilities, we created a situation in which the assumption was met.
Further, we showed that results of the SACE estimator were robust to a minor violation of the the "explainable nonrandom survival" assumption due to omission of the covariate head circumference at birth in the analysis, but not to a stronger violation due to omission of the covariate gestational age.
These results suggest that some degree of violation may be acceptable in practice. 
Omission of gestational age is quite an extreme scenario since it is probably the most obvious and important predictor of survival in preterm infants, and maybe to a slightly lesser degree, of cognitive outcome. 
Moreover, gestational age at birth is routinely available, at least in developed countries with sufficient monitoring during pregnancy.
Omission of less important predictors (due to lack of knowledge or measurement) seems more realistic.
Of note, the situation is not unlike that for methods to adjust for confounding in observational studies such as propensity score weighting, which require that no unexplained confounders exist to be unbiased (e.g. with respect to the average treatment effect), and this assumption is also untestable.
In both situations, however, subject matter expertise can be used to ensure that the most important covariates/confounders are measured and used in the analysis.
Further, the use of a cross-over trial to assess several assumptions of principal stratification methods by \cite{Qu2023} suggested that specifically the assumptions made by \cite{Hayden2005} may not be unreasonable.
Other advantages of this specific SACE estimator are the actual estimation of the SACE together with a confidence interval and its relatively convenient applicability.
Which assumptions are more or less realistic for a specific RCT should be assessed in each case.
Further, it depends on the trial design and the available data (e.g., whether longitudinal data are available), which methods can be applied.
Thus, although we used the SACE estimator of \cite{Hayden2005} in our simulation study, it is not our aim to specifically recommend this method.
A whole plethora of methods for principal stratification can be found in \cite{Lipkovich2022} and references therein, some of which can be applied to estimate the SACE.
For example, \emph{principal ignorability}, wich is a less strong assumption than \emph{explainable nonrandom survival} may be combined with the \emph{monotonicity} assumption to estimate the SACE \citep{Jo2009, Ding2017, Bornkamp2019}.

Strengths of our simulation study are that we have written and made public a simulation study protocol in advance, and that we are comparing established methods, none of which we developed ourselves.
A need for such neutral comparison studies has been identified, as studies comparing an own new method with existing methods are often overly optimistic with respect to the performance of the new method \citep{Boulesteix2013, Pawel2024}. 
A limitation of our simulation study is that we did not specifically assess situations in which the SACE ($\theta_2$) would differ from the causal effect of the treatment on the outcome in the absence of mortality ($\theta_1$). 
In fact, $\theta_2$ and $\theta_1$ were relatively similar in all scenarios, and as a consequence, the bias of the SACE estimator with regard to $\theta_2$ was only minimally smaller than the corresponding bias of the analysis using multiple imputation and vice versa.
Simulation of scenarios in which these estimands differ may require a data generating mechanism that involves interactions between treatment and covariates regarding the outcome, when some of these covariates are also related to survival.
However, these results do not mean that the SACE and the hypothetical estimand are always similar, or that the methods to estimate one can be used to estimate the other.

In clinical trials with functional outcomes truncated by death, in particular if mortality is inherent to the population studied, we believe that the SACE has a role.
In contrast to other intercurrent events, death in such populations is impossible to avoid (by trial design), which makes a hypothetical estimand less meaningful.
Similarly, the SACE has a role in trials with other intercurrent events that can not be avoided by design.
For example, the event of disease progression in oncology trials leads to truncation of other outcomes (usually secondary outcomes).
Some degree of violation of the required assumptions, in case of the SACE estimator of \cite{Hayden2005} the ``explainable nonrandom survival'' assumption, is almost certain, as it may be true for the assumptions of most methods. 
However, given that the most important covariates associated with survival are known and used in the analysis, this shortcoming may be acceptable.
Further, several estimands may be used in parallel, e.g., one for the primary analysis and another for a supplementary analysis, as suggested in other contexts \citep[e.g., the tripartite framework][]{Akacha2017, Qu2021a}.
If the SACE is used as an estimand, it is important to also consider mortality as an outcome.
In future studies we believe that it would be valuable to further explore the potential of simulation studies and crossover trials to evaluate the principal stratum approach and related assumptions.

\vspace{6pt}
\noindent {\bf{Acknowledgements}}

\noindent {\it{We thank two anonymous reviewers for their thorough reviews and very helpful comments.}}

\vspace{6pt}
\noindent {\bf{Conflict of Interest}}

\noindent {\it{The authors have declared no conflict of interest in relation to this study. SW is co-founder of Neopredix, a spin-off company of the University of Basel}}

\vspace{6pt}
\noindent {\bf{Supporting information}}

\noindent {\it{Description and application of two additional SACE methods can be found in \texttt{Supplement\_SACE\_methods.pdf}.}

\vspace{6pt}
\noindent {\bf{Data availability}}

\noindent {\it{Code to reproduce the calculations in this manuscript is available at \url{https://osf.io/bf9ht}} and as supplement to this manuscript.}

\section*{References}
\renewcommand{\bibsection}{}
\bibliographystyle{apalike}
\bibliography{other_literature,papers_citing_Hayden2005}




\end{document}